\documentclass[entropy,article,accept,moreauthors,pdftex,12pt,a4paper]{mdpi} 
\lastpage{x}
\doinum{10.3390/------}
\pubvolume{xx}
\pubyear{2014}
\history{Received: xx / Accepted: xx / Published: xx}

\usepackage{amssymb}

\Title{%
	{Statistical Power Law due to Reservoir Fluctuations}\\
	{and the Universal Thermostat Independence Principle} 
} 

\Author{%
Tam\'as S\'andor Bir\'o*, P\'eter V\'an, 
Gergely G\'abor Barnaf\"oldi and K\'aroly \"Urm\"ossy
}

\address[1]{MTA Wigner FK RMI, Konkoly-Thege M. 29-33, Budapest, Hungary }

\corres{Biro.Tamas@wigner.mta.hu, +36-1-392-2222-3388}

\abstract{
Certain fluctuations in particle number, $n$, 
at fixed total energy, $E$, lead exactly to a cut-power law distribution in 
the one-particle energy, $\omega$,
via the induced fluctuations in the  phase-space volume ratio, 
$\Omega_n(E-\omega)/\Omega_n(E)=(1-\omega/E)^n$. 
The only parameters are $1/T=\exv{\beta}=\langle n \rangle/E$
and $q=1-1/\langle n \rangle + \Delta n^2/\langle n \rangle^2$.
For the binomial distribution of $n$ one obtains $q=1-1/k$, for
the negative binomial $q=1+1/(k+1)$.
These results also represent an approximation for general particle number 
distributions in the reservoir up to second order in the canonical expansion 
$\omega \ll E$.  For  general systems the average phase-space volume ratio 
$\langle e^{S(E-\omega)}/e^{S(E)}\rangle$ to second order delivers 
$q=1-1/C+\Delta \beta^2/\langle \beta \rangle^2$
with $\beta=S^{\prime}(E)$ and $C=dE/dT$ heat capacity.
However,  $q \ne 1$ leads to  non-additivity of the Boltzmann\-Gibbs entropy, $S$.
We demonstrate that a deformed entropy, $K(S)$, can be constructed and used for demanding
additivity, i.e. $q_K=1$. This requirement leads to a second order differential equation for $K(S)$. 
Finally, the generalized  $q$-entropy formula, $K(S)=\sum p_i K(-\ln p_i)$,
contains the Tsallis, R\'enyi and 
Boltzmann\-Gibbs\-Shannon expressions as particular cases.
For diverging variance, $\Delta\beta^2$ we obtain a novel entropy formula.
}

\keyword{generalized q-entropy, fluctuations, hadronization}

\newcommand{\vs}{\vspace{3mm}}

\newcommand{\Ordo}[1]{ {\cal O}\left( {#1} \right) }

\newcommand{\be}{\begin{equation}}
\newcommand{\ee}[1]{\label{#1} \end{equation}}
\newcommand{\ba}{\begin{eqnarray}}
\newcommand{\ea}[1]{\label{#1} \end{eqnarray}}
\newcommand{\nl}{\nonumber \\}

\newcommand{\pt}[2]{ \frac{{\textrm d} #1}{{\textrm d} #2}}
\newcommand{\dt}[2]{ {{\textrm d} #1}/{{\textrm d} #2}}

\newcommand{\dif}[1]{ {\textrm d} {#1~} }
\newcommand{\ead}[1]{ ~{\textrm e}^{ #1} }

\newcommand{\exv}[1]{{\, \left\langle {#1} \right\rangle \, }}

\definecolor{Pergamen}{RGB}{235,225,200}
\definecolor{LightGray}{RGB}{235,235,230}
\definecolor{PaleBlue}{RGB}{190,210,255}
\definecolor{DarkGreen}{RGB}{0,80,20}
\definecolor{SoftRed}{RGB}{255,220,170}

\usepackage[english]{babel}
\usepackage[latin1]{inputenc}

\usepackage{times}
\usepackage[T1]{fontenc}

\usepackage{graphics}
\usepackage{graphicx}
\usepackage{fancybox}
\usepackage{bm}

\usepackage{setspace}

\usepackage{textpos}

\begin{document}


\section{Introduction}

We have been studying generalizations of the Boltzmann\-Gibbs\-Shannon (BGS) entropy 
formula\cite{RENYIorig,RenyiBook,TsallisOrigPaper,TsallisOrig2,TsallisBook}
since decades.
Our studies included the investigation of the role of multiplicative noise\cite{MultNoise,Finance},
kinetic theory\cite{NExBoltzmann,TwoGen,NExParton}, non-extensive 
equilibration\cite{PowerEquil,NExEquil,NonExtQM} and thermodynamical 
compatibility\cite{NExLimit,ZEROTH,BiroBOOK,NonAdd}, 
also with respect to infinite repetitions of abstract composition rules\cite{CompRule}.
Recently, in the quest for mechanisms explaining the occurence of a statistical power law
distribution in canonical ensembles, we emphasized the role of finite reservoir effects
in the mathematical derivation\cite{IdealGas,PowerLaw,UTI,NewEnt}.
The majority attitude to nonextensive physics is in general to start with the presentation of a
formula for the entropy and then deriving mathematical relations from it, in
order to demonstrate that the traditional requirements, like concavity, unique equilibrium
state or the Lagrange multiplier handling of secondary constraints, are fulfilled
as well as in the original 
approach\cite{WilkAPP2012,WilkPRD2013,Wilk4,Wilk6,Wilk}. 
Comparisons to experimental data then usually 
supplement the results of such investigations\cite{CooperFrye,Micro,GenTsallis}.

Our present approach reveals a different path: We start with the traditional
postulates and formulas, and then try to show why and how a ''deformation'' of the original
classical BGS entropy formula becomes unavoidable. As a by-product of such a procedure
we obtain the physical background interpretation for the parameters $T$ and $q$, 
characterizing the ubiquitous cut power law probability distribution. 
In the limit $q=1$ the BGS framework is reconstructed\cite{Almeida,Wilk,BiroBOOK,Begun1}.

As we shall demonstrate below, the common physical cause of $q \ne 1$ is the finiteness
of the physical environment, a finite heat bath\cite{Almeida,CAMPISI,BAGCI,Parvan2006,UTI}.
Whether the finite size corrections may become negligible is a case by case
problem, entangled with the physical properties of the system under study. 
Some systems, called ''non-extensive'', may behave
as finite ones in this respect even in large volumes -- since some effects
behind $q \ne 1$ depend on ratios of large quantities\cite{TsallisBook}.
To gain a feeling about the magnitude of such effects
we remind that besides the Avogadro number $\Ordo{ 10^{24}}$, considered in
classical thermodynamics of atomic matter, complex networks, like e.g. the
human brain include about the square root of this number of elements
$\Ordo{ 10^{12}}$. The internet contains approximately $10^7$ hubs
and $10^{10}$ connections. On the other hand a relativistic
heavy ion collision produces a fireball of several $\Ordo{10^3}$
new hadrons (strongly interacting particles), while in a more elementary
$pp$ collision about $\Ordo{10}$ particles are detected\cite{PHENIX2008,ALICE2013,ALICEMULT}.
Since one expects that the relative (scaled) fluctuations grow with the
decreasing number of participants, it is evident that the high energy
physics experiments are able to reveal finite reservoir effects 
quantitatively\cite{Begun1,Begun2,KOCH,BEGUN,Wilk4,Wilk6,Gorenstein,Gorenstein2,BegunGoren,PowerLaw,UTI}.

In this paper we seek answer to the following two questions:
i) What is the physics behind $q \ne 1$
and ii) what $K(S)$ deformation of the entropy $S$ is necessary to achieve $q_K=1$?
We note that $q=1$ signalizes an additive composition rule, so the second
question is equivalent for seeking an additive ({\em ''K-additive''}) description
in case of non-negligible finite size corrections on the classical 
thermodynamics\cite{UTI,NonAdd,ZEROTH}.





\section{Finite Heat Bath and Fluctuation Effects}

In this section we review the traditional approach
to the thermodynamical statistical weight assuming
a uniform phase-space distribution of microstates\cite{MaBook1985}.
At the beginning we present a very simple model of
particle production, where the total energy, $E$, is fixed
(in experiments  $\Delta E/E \lesssim 10^{-3}$),
but the number of produced particles, $n$, fluctuates appreciably.
Its distribution will be considered first in terms of
the simplest possible assumptions about combining occupied
and unoccupied phase-space cells in a 
{\em finite observed section of the available total phase-space}.
Following this analysis more general $n$ distributions and finally
a general heat bath, described by its equation of state, $S(E)$,
is considered. During this chain of models we seek answer for
the question:

\subsection{What is the physics behind the parameter $q$ ?}


Our starting point is an ideal gas in a finite phase-space\cite{IdealGas,Micro,MaBook1985,Parvan2006}.
We describe the microcanonical statistical weight for having
a one-particle energy, $\omega$, out of total energy, $E$. 
In a one-dimensional relativistic jet it is distributed
according to the ratio of corresponding phase-space volumes as
\be
P_1(\omega) = 
\frac{\Omega_1(\omega) \, \Omega_n(E-\omega)}{\Omega_{n+1}(E)} =
\: \mathrm{\rho}(\omega) \: \cdot \frac{\left(E-\omega\right)^n}{E^n}
\ee{ONEPARTICLE}
Here $\Omega_{n+1}(E)$ is the total phase-space, while $\Omega_n(E-\omega)$ is the
phase-space for the reservoir, missing one particle with energy $\omega$.
The number of particles,  $n$, itself has a distribution
(based on the physical model of the reservoir and on the event by event detection of the spectra).

We consider ideal reservoirs with a (negative) binomial $n$-distribution, obtained
from the following simple argumentation.
We distribute {$n$ particles among $k$ cells:} 
bosons in $\binom{n+k}{n}$ ways, fermions in $\binom{k}{n}$ ways.
By observation we detect a subspace $(n,k)$ out of a bigger $(N,K)$ reservoir.
The limit $K\rightarrow\infty$, $N\rightarrow\infty$ with a 
fixed average occupancy $f=N/K$, constitutes the traditional canonical limit.
However, we keep here several finite size factors. We obtain
\be
B_{n,k}(f) := \lim_{K\rightarrow\infty}\limits 
\frac{\binom{n+k}{n}\binom{N-n+K-k}{N-n}}{\binom{N+K+1}{N}}
= {{ {\binom{n+k}{n} \, f^n \, (1+f)^{-n-k-1}}} }
\ee{NBD_DEF}
for bosons and
\be
F_{n,k}(f) := \lim_{K\rightarrow\infty}\limits 
\frac{\binom{k}{n}\binom{K-k}{N-n}}{\binom{K}{N}}
= {{ {\binom{k}{n} \, f^n \, (1-f)^{k-n}}} }
\ee{BD_DEF}
for fermions.

Since most hadrons produced in high-energy experiments are pions, which are bosons,
we consider first the bosonic reservoir described by $B_{n,k}(f)$. 
The average statistical weight factor, $w_E(\omega)$, with
fixed $E$ and the negative binomial distribution (NBD) of $n$ becomes
\be
w^{\rm NBD}_E(\omega) = \sum_{n=0}^{\infty}\limits \left(1-\frac{\omega}{E} \right)^n B_{n,k}(f) =
 \left[ (1+f)-f\left(1-\frac{\omega}{E}\right) \right]^{-k-1} =
 \left(1+ f\frac{\omega}{E} \right)^{-k-1}
\ee{NBD_AVER}
Note that $\exv{n}=(k+1)f$ for NBD. Then with the notation
$T=E/\exv{n}$ and $q-1=\frac{1}{k+1}$ we get
\be
w^{\rm NBD}_E(\omega) = \left(1 + (q-1) \frac{\omega}{T} \right)^{-\frac{1}{q-1}}.
\ee{qBGTSALLIS}
{This is {\em exactly} a $q>1$ Tsallis\-Pareto distribution}.
The opposite correspondence, namely that an assumed Tsallis\-Pareto distribution
leads to an NBD multiplicity distribution, has been pointed out by Wilk and Wlodarczyk
\cite{Wilk,Wilk3,Wilk6}.
Experimental NBD distributions of total charged hadron multiplicites stemming from 
Au + Au collisons at $\sqrt{s}_{NN}=62$ and $200$ GeV 
can be inspected e.g. in~\cite{PHENIX2008}.
Characteristically $k \approx 10 - 20$, therefore $q \approx 1.05 - 1.10$
\cite{PHENIX2008,ALICE2013,ALICEMULT}.

For a fermionic reservoir $n$ is distributed according to 
the Bernoulli distribution (BD). The average phase-space volume ratio
becomes
 \be
 w^{\rm BD}_E(\omega) =  \sum_{n=0}^{\infty}\limits \left(1-\frac{\omega}{E} \right)^n F_{n,k}(f) =
 \left[ (1-f)+f\left(1-\frac{\omega}{E}\right) \right]^{k} =
 \left(1- f\frac{\omega}{E} \right)^{k}
 \ee{BD_AVER}
Note that $\exv{n}=kf$ for BD. Then with $T=E/\exv{n}$ and $q-1=-\frac{1}{k}$
we obtain  {\em exactly} a $q<1$ Tsallis\-Pareto distribution,
$$
w^{\rm BD}_E(\omega) = \left(1 + (q-1) \frac{\omega}{T} \right)^{-\frac{1}{q-1}}.
$$
It is enlightening to consider the Boltzmann\-Gibbs limit of the above.
In case of low occupancy in the phase-space, $k \gg n$ and both the BD and NBD
distributions approach a Poissonian:
\be
\Pi_n \: = \: \frac{\exv{n}^n}{n!} \ead{-\exv{n}}  \quad \mathrm{with} \quad    
\exv{n} = k~ \frac{f}{1\pm f} \quad \mathrm{fixed}. 
\ee{POISSON}
The resulting statistical factor is {\em exactly} the Boltzmann\-Gibbs exponential
with $T=E/\exv{n}$,
\be
w^{\rm BG}_E(\omega) = \sum_{n=0}^{\infty}\limits \left(1 - \frac{\omega}{E}\right)^n \Pi_n(\exv{n}) 
= \ead{\left(1-\omega/E \right)\exv{n}} \ead{-\exv{n}} 
= \ead{-\exv{n}\omega/E} = \ead{-\omega/T}.
\ee{POIS_AVER}
In all of the three above cases the parameter $T$ is defined by the
(one-dimensional, extreme relativistic) equipartition,
and $q$ is related to the scaled variance of the produced particle number:
\be
  T = \frac{E}{\exv{n}},   \quad \mathrm{and} \quad 
  q = 1- \frac{1}{\exv{n}} + \frac{\Delta n^2}{\exv{n}^2}. 
\ee{AVERG_ALL}
For general $n$-fluctuations, $P_n$, the above result also applies,
albeit {\em only as an approximation}.
In the philosophy of the canonical approach we expand our formulas for small $\omega \ll E$.
The Tsallis\-Pareto distribution as an approximation reads as
\be
\left(1+(q-1)\frac{\omega}{T} \right)^{-\frac{1}{q-1}} =
1-\frac{\omega}{T} + q \frac{\omega^2}{2T^2} - \ldots
\ee{TSALLIS_EXPAND}
The ideal reservoir phase-space ratio up to second order in this limit results in
\be
w_E(\omega) = \exv{\left(1-\frac{\omega}{E}\right)^n} =
1 - \exv{n}\frac{\omega}{E} + \exv{n(n-1)}\frac{\omega^2}{2E^2} - \ldots
\ee{RESERV_EXPAND}
Comparing the corresponding coefficients one concludes that 
eq.~(\ref{AVERG_ALL}) as an approximation 
holds for a general $n$-distribution.


Finally we deal with a general environment, given by its equation of state, $S(E)$.
In the expansion for small $\omega \ll E$ the phase-space volume ratio becomes
\be
w_E(\omega) = \exv{\frac{\Omega_n(E-\omega)}{\Omega_n(E)}} =
\exv{e^{S(E-\omega)-S(E)}} = \exv{e^{-\omega S^{\prime}(E)+\omega^2 S^{\prime\prime}(E)/2 - \ldots}}
\nonumber
\ee{SMAL_OMEGA}
\be
= 1 - \omega \exv{S^{\prime}(E)} + \frac{\omega^2}{2} \exv{S^{\prime}(E)^2+S^{\prime\prime}(E)} - \ldots
\ee{COMPLEMENT_PHASE_SPACE}
Comparing it with the expansion of the Tsallis\-Pareto distribution, 
eq.~(\ref{TSALLIS_EXPAND}), one concludes
\be
\frac{1}{T} = \exv{\beta} = \exv{S^{\prime}(E)}, \qquad 
q = 1 - \frac{1}{C} + \frac{\Delta \beta^2}{\exv{\beta}^2}.
\ee{INTERPRET}
This is the final interpretation of the parameters $T$ and $q$ for a general reservoir.
Note that due to $\exv{S^{\prime\prime}(E)}=-1/CT^2$, our result is 
expressed via the heat capacity of the reservoir, defined as $1/C = \dt{T}{E}$.
In general we have {opposite} sign contributions 
from $\exv{S^{\prime \: 2}}-\exv{S^{\prime}}^2$ and from $\exv{S^{\prime\prime}}$. 
In the light of this result one realizes that
\begin{itemize}
\item $q>1$ and $q<1$ are both possible,
\item for any relative variance $\Delta \beta /\exv{\beta} = 1/\sqrt{C}$ it is exactly $q=1$, 
\item and for $E \propto n/\beta = \text{const}$  
	we have $\Delta \beta/\exv{\beta} = \Delta n / \exv{n}$. 
\end{itemize}
In this way the $n$-fluctuations represent a particular case of the more general
reservoir fluctuations.

At the end of this section we sketch the relation of our approach to
{\em superstatistics}\cite{SuperStat,Beck2,Wilk5,Beck:dyn,Kodama:dyn}. 
In its original formulation superstatistics dealt with
fluctuations of the Lagrange multiplier $\beta$. Demanding that we describe
the same non-exponential statistics, only in two different ways, one arrives at
the relation
\be
\int e^{-\beta\omega} \, \gamma(\beta) \, \dif{\beta} \: 
= \: \sum_n P_{\!n}(E) \, \left(1-\frac{\omega}{E}\right)^n
\ee{SUPER1}
Note that $ e^{-\beta\omega} = e^{\left(1-\frac{\omega}{E}\right) \, \beta E } \, e^{-\beta E} $.
Using now the Taylor series of the first exponential one obtains 
\be
P_{\!n}(E) \, = \int \frac{(\beta E)^n}{n!} \, e^{-\beta E} \, \gamma(\beta) \, \dif{\beta}.
\ee{SUPER2}
The converting factor is a Poissonian with the parameter $\overline{n}= \beta E$.
Inverting the above procedure one seeks for a superstatistics from the $n$-distribution. 
Applying the correspondence eq.~(\ref{SUPER1}) for $\omega = E$:
\be
\int e^{-\beta E} \, \gamma(\beta) \, \dif{\beta} \: = \: P_0(E).
\ee{CORRESP1}
Inverse Laplace transformation then, in principle, delivers the superstatistical factor
\be
\gamma(\beta) = {\cal L}^{-1} \left[P_0(E) \right].
\ee{SUPERFACTOR}
Expanding for small $\omega$, however, one gets
$\exv{\beta} = {\exv{n}}/{E}$ and $\exv{\beta^2} = {\exv{n(n-1)}}/{E^2}$, 
leading to
\be
q = 1 + \frac{\Delta \beta^2}{\exv{\beta}^2} = 1 + \frac{\Delta n^2}{\exv{n}^2} - \frac{1}{\exv{n}}.
\ee{BAJVAN}
One immediately realizes that for some $n$-distributions, alike the BD,  
$\Delta \beta^2$ would have to be negative. It is impossible. This problem is
also reflected in the fact that
there is no guarantee that an inverse Laplace transformation results in an
overall positive function. In this way the superstatistics due to $n$-fluctuations, $P_n(E)$,
seems to be more general, than the approach with solely a $\beta$-distribution, $\gamma(\beta)$.
In particular a statistical $\beta$-distribution cannot ever match a $q<1$ result.


\section{Deformation of the entropy}

Once we understood how and why finite reservoir effects lead to $q \ne 1$,
and emerging from this to a non-exponential statistical weight, the need
for mending this salient feature arises. Generalizing the Boltzmann\-Gibbs exponential
to another formula, containing finite reservoir corrections, also abandons the
remarkable basic property of the exponential: the additivity of the arguments
by the product. Since this property connected the dynamical independence
(energy additivity) with the statistical independence (probability factorization
or equivalently entropy additivity), its missing is a severe conundrum.

In this section we show, that if the original logarithmic definition due to
Boltzmann or equivalently its exponential inverse due to Einstein,
postulating the phase-space volume to be proportional to the exponential
of the entropy, fails to some degree, then one may search for another
expression of the entropy, $K(S)$, in order to restore  ''K-additivity''.
We comprise our quest into the simple question:
{If $S$ leads to $q \ne 1$, what $K(S)$ achieves $q_K=1$?}

\subsection{The additive entropy $K(S)$}

We call ''deformed entropy'' the quantity $K(S)$, being additive while $S$ was non-additive.
In the basic postulate we use $K(S)$ instead of $S$ 
in the exponential in order to gain more flexibility for handling the subleading term in 
the $\omega \ll E$ expansion discussed above and shown to interpret the parameter $q$.
In this way we consider
\ba
w^{\rm K}_E(\omega) = \exv{e^{K(S(E-\omega))-K(S(E))}}  \, = \, 1- \omega \exv{\pt{}{E} K(S(E))} 
\nl
 \, + \, \frac{\omega^2}{2} \exv{\pt{^2}{E^2} K(S(E)) + \left(\pt{}{E} K(S(E))\right)^2} 
\, + \, \ldots
 \ea{DEFORMED_ENTROPY_STATISTICAL_WEIGHT}
Note that
$\pt{}{E} K(S(E)) = K^{\prime} S^{\prime}$ and
$ \pt{^2}{E^2} K(S(E)) = K^{\prime\prime} S^{\prime \, 2} + K^{\prime} S^{\prime\prime}$.
Now we compare this expression with the Tsallis\-Pareto power-law.
Using previous average notations and assuming that $K(S)$ is independent of the
reservoir fluctuations (a certain {\em universality}) one obtains:
\be
\frac{1}{T_K} = K^{\prime} \frac{1}{T},
\qquad
\frac{q_K}{T_K^2} = \left(K^{\prime\prime}+K^{\prime \, 2} \right) \frac{1}{T^2} 
\left(1+\frac{\Delta \beta^2}{\exv{\beta}^2} \right) - K^{\prime} \frac{1}{CT^2}.
\ee{KS_TASSILS_PAREMETERS}
By choosing a particular $K(S)$ {\em we shall  manipulate $q_K$}.
In order to simplify the differential equation posed on $K(S)$ by requiring a
given value for $q_K$ we introduce  the notations {$F = 1 / K^{\prime} = T_K / T$}  and 
{$\Delta \beta^2/\exv{\beta}^2 = \lambda / C $}.
Then the {$q_K$ parameter for the $K(S)$  entropy} is expressed as 
\be
q_K \: = \: \left(1+\frac{\lambda}{C} \right) \, \left(1-F^{\prime} \right) \: - \: \frac{1}{C} \, F.
\ee{qK_WITH_F}
Re-arranged this represents a very simple differential equation with $q=1+\lambda/C-1/C$:
\be
(\lambda + C) F^{\prime} + F = \lambda + C(1-q_K) = 1 + C (q-q_K).
\ee{SIMPLE_DEQ}
From this form it is easy to realize that two special choices are worth to be considered:
$q_K=q$ and $q_K=1$. Since we seek for entropy deformations with the property
$K(0)=0$ and $K^{\prime}(0)=1$, one fixes the condition $F(0)=1$.
In this case the only solution for $q_K=q$ is $F=1, K(S)=S$. It is obvious that the
other choice, $q_K=1$, is the only purposeful deformation for reaching 
K-additivity\cite{ZEROTH,NonAdd}.
Eq.~(\ref{SIMPLE_DEQ})  becomes then easily solvable.
We call this form of the $q_K=1$ requirement the  {\em ''Additivity Restoration Condition''}
(ARC): 
\be
\fbox{ $\left(\lambda + C \right) \, F^{\prime} \, + \, F \: = \: \lambda.$  }
\ee{F_SOLVABLE}

\vs
\subsection{Classification by Fluctuation Models}

$q_K=1$ also means a re-exponentialization of the $\omega$-expansion of
the statistical weight based on the deformed entropy phase-space, $w^{\rm K}_E(\omega)$.
In this way the effective equilibrium condition, the common temperature,
least depends on the one-particle subsystem energy, $\omega$. In earlier
publications we called this the ''Universal Thermostat Independence'' (UTI)
principle\cite{UTI}.

Now we explore the solutions of the ARC equation (\ref{F_SOLVABLE})
under different assumptions about the heat capacity and the reservoir
fluctuations. In the simplest case we do not consider
reservoir fluctuations at all, we put $\Delta \beta^2=0$ and therefore
$\lambda=0$.  Applying our previous general result for this value we have to solve 
\be
F^{\prime} \, + \, \frac{1}{C} \, F \: = \: 0.
\ee{qK}
Replacing back the definition $F=1/K^{\prime}$, one arrives at the original 
UTI equation~\cite{UTI}:
\be
\frac{K^{\prime\prime}}{K^{\prime}} = \frac{1}{C}.
\ee{UTI_EQUATION}
For ideal gas $C=1/(1-q)$ is constant, 
and the solution of eq.~(\ref{UTI_EQUATION}) with \, $K(0)=0, \: K^{\prime}(0)=1$ \, 
delivers\cite{IdealGas,Almeida,BAGCI}
\be
K(S) = C \left(e^{S/C}-1 \right).
\ee{UTI_SOLUTION}
From this result one arrives upon using $K(S)=\sum_i p_i K(-\ln p_i)$
at the statistical entropy formulas of  {Tsallis and R\'enyi:}
\cite{TsallisOrigPaper,TsallisOrig2,TsallisBook,RENYIorig,RenyiBook}
\be
K(S) = \frac{1}{1-q} \sum_i \left(p_i^{q}-p_i\right), \qquad
S = \frac{1}{1-q} \ln \sum_i p_i^{q}.
\ee{RENYI_TSALLIS}
Next we obtain the deformed entropy formula with $C$ and $\lambda$ constant.
Using eq.~(\ref{F_SOLVABLE}) one obtains the general differential equation
\be
\lambda K^{\prime \, 2} -  K^{\prime}
+ C_{\Delta} K^{\prime\prime} = 0
\ee{qK_ONE_DIFF_EQ}
with $C_{\Delta}=C+\lambda$.  Its first integral, 
\be
K^{\prime}(S) = \frac{1}{(1-\lambda)e^{-S/C_{\Delta}}+\lambda} 
\ee{KPRIME_FOR_qK_ONE}
and second integral, 
\be
K(S) = \frac{C_{\Delta}}{\lambda}  \,  \ln \left(  1-\lambda + \lambda e^{S/C_{\Delta}} \right),
\ee{K_FOR_qK_ONE}
represent the optimal deformation of the entropy formula in this case.
With the above result (\ref{K_FOR_qK_ONE}) the $K(S)$-additive  composition rule, 
$ K(S_{12}) = K(S_1)+K(S_2)$, is equivalent to
\be
h(S_{12}) = h(S_1) + h(S_2) + \frac{\lambda}{C_{\Delta}} h(S_1) h(S_2)
\ee{qK_COMPOSITION}
with
\be
h(S) = C_{\Delta} \left( e^{S/C_{\Delta}}-1 \right).
\ee{DEFINE_hS}
This is a combination of the {ideal gas} entropy-deformation, $h(S)$ and
an {original Tsallis} composition law\cite{TsallisBook,Abe}
with $q-1=\lambda/C_{\Delta}$.
Using the auxiliary function, $h_C(S)=C(e^{S/C}-1)$, we have $h_{\infty}(S)=S$ and
the entropy deformation function can also be written as
\be
K_{\lambda}(S) \, = \, h^{-1}_{C_{\Delta}/\lambda}\left( h_{C_{\Delta}}(S)\right).
\ee{K_DEFORM}
\begin{itemize}
\item
{ For ${\lambda=1}$ it is {obviously} ${K_1(S)=S}$.} 
This is the Gaussian fluctuation model, considered in several textbooks, and also
believed to lead to the smallest physically possible variance due to
a ''thermodynamical uncertainty'' principle\cite{Uffink1,Lavenda,Uffink2}. 
Since $\beta=S^{\prime}(E)$, the variances are related as 
$\Delta\beta=\left|S^{\prime\prime}(E) \right| \Delta E = \Delta E/CT^2$.
Then from $\Delta\beta \cdot \Delta E \ge 1$ it follow $\Delta E \ge T\sqrt{C}$
and $\Delta\beta \ge 1/T\sqrt{C}$. A straightforward consequence of this is
$\lambda/C=\Delta \beta^2/\exv{\beta}^2 \ge 1/C$ and therefore $\lambda \ge 1$.
We note, that if this ''uncertainty'' principle were
correct, then only $q>1$ canonical distributions of $\omega$ would exist in Nature.

\item
For no fluctuations ${\lambda = 0}$ and we get ${K_0(S)=h_C(S)}$.
We regain the Tsallis and R\'enyi formulas presented above in eq.~(\ref{RENYI_TSALLIS}).

\item
It is also very intriguing to inspect the following particular limit: 
$C \rightarrow \infty, \lambda\rightarrow\infty$ but
$\lambda/C_{\Delta} \rightarrow \tilde{q}-1$ finite. In this
non-extensive limit the fluctuations are much larger than the normal Gaussian ones,
and we obtain a nontrivial entropy deformation:
\be
K_{NE}(S) = h^{-1}_{1/(\tilde{q}-1)}\left( h_{\infty}(S) \right) = 
\frac{1}{\tilde{q}-1} \ln \left(1+(\tilde{q}-1)S \right).
\ee{NONEXT_KOFS}
The K-additivity, $K(S_{12})=K(S_1)+K(S_2)$, in this case leads to the
non-additivity formula $S_{12}=S_1+S_2+(\tilde{q}-1)S_1S_2$, 
-- investigated formerly in depth by Tsallis and 
Abe\cite{TsallisBook,Abe,Abe2,Abe3,AbeRaj,AbeRaj1,AbeRaj2,AbeBagci:qexp}.
\end{itemize}
In the finite heat capacity, finite temperature variance case
we arrive at a {\em Generalized Tsallis Formula}
based on $K(S)=\sum_i p_i \, K(-\ln p_i)$: 
\be
K_{\lambda}(S) = \frac{C_{\Delta}}{\lambda} \, \sum_i p_i \,
 \ln \left(1-\lambda + \lambda p_i^{-1/C_{\Delta}} \right).
 \ee{GENERAL_TSALLIS_qK_ONE}
\begin{itemize}
\item
For normal fluctuations
$K_1(S)=-\sum_i p_i \, \ln p_i $ is {exactly the Boltzmann entropy}.

\item
Without fluctuations
$K_0(S)= C \sum_i \left(p_i^{1-1/C}-p_i \right)$  is the {Tsallis entropy}
with $q=1-1/C$ and $S$ is the corresponding R\'enyi entropy.

\item
Finally considering extreme large fluctuations and a finite heat capacity, $C(S)$
which however may be an arbitrary function of the total entropy, $S$, we obtain
the non-extensive result eq.~(\ref{NONEXT_KOFS}) with $\tilde{q}=2$:
\be
K_{\infty}(S) = \ln \left( 1 + S \right) = \sum_i p_i \ln \left( 1 - \ln p_i \right).
\ee{KS_qK_ONE_LAMBDA_INFTY}
The canonical $p_i$ distribution maximizing this parameterless deformed entropy
is a Lambert W function, it shows tails like the 
{Gompertz distribution}\cite{Gompertz,Casey,Apostol},
known from extreme value statistics and nonequlibrium growth models for
demography and tumors. 
\end{itemize}

\section{Conclusion and Outlook}

In conclusion we have shown that in terms of traditional phase-space models
the statistical cut power-law behavior can be interpreted as being primarily
a particle number fluctuation effect during hadronization in high energy
collisions. The $q>1$ and $q<1$ Tsallis\-Pareto distributions are exact
for NBD and BD distributions of the particle number, respectively, in a one-dimensional
phase-space characteristic for high energy jets. The Boltzmann\-Gibbs exponential
weight factor is restored for the common limiting case of these distributions,
for the Poissonian, leading to $q=1$.

For general particle number distributions with fixed energy the Tsallis\-Pareto
cut power-law is only an approximation to subleading order in the expansion
for small individual energy, $\omega \ll E$. We obtained and interpereted the parameters
$T$ and $q$ by comparing coefficients of the respective expansions and concluded
that $T=E/\exv{n}$ is an equipartition temperature, while $q=1+\Delta n^2/\exv{n}^2-1/\exv{n}$
reflects both the particle number variance and due to its expectation value the
size of the reservoir. This formula also explains why both $q>1$ and $q<1$
cases can be observed in natural phenomena.

Further generalization towards the thermodynamical treatment considers the
reservoir environment described by a simplified equation of state, $S(E)$.
Repeating the above described approximations one concludes that 
$1/T=\exv{\beta}=\exv{S^{\prime}(E)}$, i.e. the parameter $T$ also plays the role
of a thermodynamical temperature. The parameter $q$ is again related both to
the size (total heat capacity, $C$) of the reservoir and to the variance of
the fluctuating quantity $\beta=S^{\prime}(E)$. The general formula follows
the structure obtained in the high energy model, $q=1+\Delta\beta^2/\exv{\beta}^2-1/C$,
with $1/C=\dt{T}{E}=-T^2\exv{S^{\prime\prime}(E)}$.

It is, however, known for long that the cut power-law does not follow the
product rule, as the Boltzmann\-Gibbs exponential does, for additive energy.
The root of this behavior is the non-additivity of the Boltzmannian entropy, $S$,
for finite and fluctuating reservoirs. 
$S(E_1+E_2) \ne S(E_1)+S(E_2)$ for $q \ne 1$
is a weakness of the classical thermodynamics which has to be cured.
Our approach here was to look for a function, $K(S)$, which restores additivity
by leading to $q_K=1$.
This requirement for such a function concludes in the additivity restoring
condition, ARC, in a differential equation satisfied by $K(S)$.
Finally the usual canonical treatment must then be based on the additivity
of $K(S)$, applied to an ensemble of configurations, which in turn
provides the general formula $K(S)=\sum p_i K(-\ln p_i)$ 
(cf eq.~\ref{KS_qK_ONE_LAMBDA_INFTY} and \cite{IdealGas}).

The Boltzmann\-Gibbs\-Shannon formula is restored for $q=1$ (when also $K(S)=S$
is the only solution), in particular for the traditional Gaussian approach to fluctuations
when $\Delta \beta/\exv{\beta}=1/\sqrt{C}$ is taken for granted.
When the fluctuations are negligible, the Tsallis entropy formula arises
for $K(S)$ and the corresponding R\'enyi formula for $S$ with $q=1-1/C$.
In the extreme large fluctuation limit a new, up to now not considered entropy
-- probability formula arises.

These initial results are encouraging for further pursuit of such a theoretical
approach. The research of large systems, where $\lambda=C\Delta\beta^2/\exv{\beta}^2 \gg C \gg1$
with a finite limit for $\lambda/C$, shall deal with genuine non-additivity
of the Boltzmann entropy. The physical modelling of the reservoir environment,
in particular with emphasis on the variable number of particles relevant for
high energy physics, leads to more complex descriptions than presented here:
a dependence like $C(S)$ and $\lambda(S)$ can be quite common. In such cases
the ARC differential equation leads to further entropy formulas.
Our approach provides a procedure to find the optimal entropy -- probability
relation form the viewpoint of the non-additive composition of two (or gradually more)
subsystems. Also the superstatistics, originally conceptualized as a $\beta$-distribution
behind non-Gibbsean factors in the statistics, may be extended to studies
considering physical systems which cannot be described simply by an overall
positive weight factor $\gamma(\beta)$ under an integral.


\acknowledgements{Acknowledgements}

This work was supported by the Hungarian National Research Fund
OTKA (Grants K 104260, NK 106119) 
and by a bilateral Chinese\-Hungarian grant NIH TET\_12\_CN-1-2012-0016. 
G.G.Barnaf\"oldi thanks the support in form of  
the J\'anos Bolyai Research Scholarship of the Hungarian
Academy of Sciences.


\authorcontributions{Author Contributions}

The content of this article was presented to a great part by T.~S.~Bir\'o
at the {\em Sigma Phi 2014} conference at Rhodes, Greece,
in an invited talk.


\conflictofinterests{Conflicts of Interest}

The authors declare no conflict of interest. 

\bibliographystyle{mdpi}
\makeatletter
\renewcommand\@biblabel[1]{#1. }
\makeatother

\end{document}